# A Study of the B/C Ratio Between 10 MeV/nuc and 1 TeV/nuc in Cosmic Rays Using New Voyager and AMS-2 Data and a Comparison with the Predictions of Leaky Box Propagation Models

W.R. Webber and T.L. Villa

New Mexico State University, Department of Astronomy, Las Cruces, NM, 88003, USA

**ABSTRACT**

This paper seeks to find an explanation of the galactic cosmic ray B/C ratio newly measured in cosmic rays between ~10 MeV/nuc and 1 TeV/nuc. Voyager measurements of this ratio at low energies and AMS-2 measurements at high energies are used in this study. These measurements both considerably exceed at both low and high energies the ratio predicted using a simple Leaky Box Model for propagation of cosmic rays in the galaxy. Between 1-70 GeV/nuc, however, this same model provides an excellent fit (within $\pm$2-3%) to the new AMS-2 measurements using an escape length ~$P^{-0.45}$. This would imply a diffusion coefficient ~$P^{0.45}$, very close to the Kraichian cascade value of 0.50 for the exponent. Extending this same diffusion dependence to high energies, still in a LBM, along with a truncation of short path lengths in the galaxy will predict a B/C ratio of 4.5% at ~800 GeV/nuc which is very close to the AMS-2 measurement at that energy. This would indicate that the amount of material traversed near the source in a Nested LBM, for example, is less than about 0.5 g/cm$^2$ at these energies since this process, with additional matter near the sources, would increase the B/C ratio. At low energies, however, the B/C ratio of 14% $\pm$3% at ~10 MeV/nuc measured by Voyager is more difficult to explain. The same parameters used in a LBM that fit the high energy B/C measurements predicts a B/C ratio of only ~4% at 10 MeV/nuc where the path length is about 1.5 g/cm$^2$. If the cosmic rays have traversed ~10 g/cm$^2$ of material in the galaxy at low energies the prediction for B/C is still only 8%, which is between 2.0-3.0 sigma below the measurement and its errors.

**Introduction**

New Voyager measurements of B and C nuclei are now available beyond the heliopause down to below 10 MeV/nuc (Cummings, et al., 2016). AMS-2 measurements of these same nuclei have recently been reported up to nearly 1 TeV (Oliva, et al., AMS-CERN, 2015). From these and other data the B/C ratio can now be obtained over a range of 5 orders of magnitude. This ratio is of crucial importance for understanding both the origin and propagation of galactic cosmic rays since it contains a purely secondary nucleus, Boron, and a primary (source) nucleus, Carbon.

The measured B/C ratio at Voyager shows an excess at low energies above the calculations of a simple Leaky Box propagation model as will be described in this paper. However between about 1 and 70 GeV/nuc (3-150 GV) the calculations in this paper and measurements of AMS-2 of this ratio are in excellent agreement to within a few percent for a dependence of cosmic ray escape length which is $\sim P^{-0.45}$. This implies a diffusion coefficient $\sim P^{0.45 \pm 0.05}$, quite close to, but not exactly, a Kraichian cascade (Kraichian, 1965).

As a result of the measured excesses in the B/C ratio by Voyager and AMS-2 above those obtained using a simple LBM calculation, we have considered modified LBM's with a significant amount of matter traversed near the sources at both low and high energies, essentially a Nested LBM (see Cowsik, Burch and Madziwa-Nussinov, 2014) as well as modified LBM which include a large path length in the galaxy at low energies and also a truncation at small path lengths of the exponential path length distribution in a simple LBM (see Shapiro and Silberberg, 1970; Garcia-Munoz, et al., 1987; Webber, 1993).

The goal of this paper is to examine the B/C ratio over an extended energy range using the new Voyager data at low energies and the AMS-2 data at high energies within the framework of the Leaky Box Model and with modifications that may be successful in explaining both the low and high energy data.

**The Data**

Figure 1 shows the B/C nuclei ratio in galactic cosmic rays from ~10 MeV/nuc to 1 TeV/nuc as a function of energy. This ratio is one of the most sensitive indicators we have of the amount of matter that primary cosmic rays have traversed since their acceleration, since B is a purely secondary nuclei with a zero source abundance. Shown in the figure are the AMS-2 data (Oliva, et, al., AMS-CERN, 2015) at energies above ~0.5 GeV/nuc, which are made at significant solar modulation levels and the Voyager data at lower energies (Cummings, et al., 2016), made at a near zero solar modulation level beyond the heliopause and also the ACE-CRS data at the Earth at a solar modulation level ~250 MV (Lave, et al., 2013). Also shown (lower solid curve) is the calculated B/C ratio made using a Leaky Box model with an escape length, $\lambda = 22.3 \beta P^{-0.45}$ above 1 GV and $33.9 \beta^{3/2}$ below 1 GV (see Webber, 2015a). These calculations use the latest available cross sections for the production of B (see Webber, Higbie and McCracken, 2007 and Webber, 2015b) and are also based on a fit to lower energy H and He spectra measured at Voyager (Webber and Higbie, 2013). The B/C ratio is calculated at energies below ~1 GeV/nuc for solar modulation levels ~1000 and 250 MV appropriate to the AMS-2 and ACE data, respectively, and is shown as blue lines.

Also shown in Figure 1 as a dashed line is the B/C ratio at both low and high energies calculated with the same propagation parameters as used in the standard LBM but using a truncated PLD at short path lengths with a truncation parameter defined as = 0.12. This value of 0.12 for the truncation parameter has been used to fit the Voyager measured low energy part of the spectra for primary C → Fe nuclei as described in Webber, 2016. This truncation, which is described in great detail in the Garcia-Munoz, et al., 1987, paper, increases the B/C ratio mostly at both low and high energies where the mean path length is small (less than 4 $g/cm^2$) while at the same time it decreases the intensities of primary nuclei < 100 MeV/nuc as compared with a standard LBM as discussed in Webber, 2016. So it will increase the B/C ratio significantly for a given set of propagation parameters as is shown by the dashed lines.

The agreement to within a few percent between these LBM predictions and the new AMS-2 data at energies between 1-70 GeV/nuc where the solar modulation becomes small, as shown in Figure 1, underscores the precision with which a LBM can be used to describe the cosmic ray observations and the matter path length traversed in the galaxy by the cosmic ray source nuclei in this energy range. The matter path length in $g/cm^2$ used for the various propagation models used in this paper is illustrated in Figure 2.

It is seen in Figures 1 and 2 that a $P^{-0.45}$ dependence describes the observed rigidity dependence of the path length above a few GeV/nuc and up to ~100 GeV/nuc as determined from the B/C ratio. At 100 GeV/nuc this mean path length is ~2.50 g/cm$^2$. If this same dependence continues up to 800 GeV/nuc (the maximum energy of the AMS-2 measurements) the mean path length for cosmic ray primary nuclei at this energy would be ~1.00 g/cm$^2$.

We note in Figure 2 that for a break in the path length dependence at 1 GV the mean path length decreases at low energies and at about 25 MeV/nuc it is also 2.50 g/cm$^2$, the same as at 100 GeV/nuc. This rapidly decreasing path length, below a rigidity $P_0$, which is about 1.0 GV, is supported by the measurements of the electron spectrum between 5-60 MeV at Voyager (Webber and Higbie, 2013; Webber, 2015a). This implies that the production of secondary nuclei such as B in interstellar space should rapidly become smaller at energies below ~120 MeV/nuc because of the smaller path length. These smaller path lengths are caused by the increasingly large diffusion coefficients at both high and low energies and, as a consequence, a rapid escape of cosmic rays from the galaxy and a shorter cosmic ray lifetime.

The value for the break rigidity is uncertain but could be between 0.316-1.0 GV, based on fits to the Voyager electron data between 3-60 MeV. So we show in Figure 2 the path length for these other possibilities. At 10 MeV/nuc which is the lowest energy point at which the B/C ratio is measured, the path length is ~1.2 g/cm$^2$ for a break at 1GV. For a break at 0.562 GV it is about 2.6 g/cm$^2$ and for a break at 0.316 GV it is about 6.0 g/cm$^2$. This factor of ~5.0 increase in the mean path length at 10 MeV/nuc will result in an increase of a factor ~2.0 in the B/C ratio at this energy as described below.

**Model Description and Calculation-Modifications to the Simple LBM**

**A - Changes in the Mean Path Length at Low Energy**

The rigidity dependence of the mean path length below ~1.0 GV needs to be re-examined in order to attempt to explain the high values of the observed B/C ratio at the lowest energy. The dependence of this path length has been taken to be $\lambda = 33.9\ \beta^{3/2}$ below 1.0 GV as noted above. Other possibilities that we have calculated in a separate paper on B and N nuclei (Webber, 2015b) include $\lambda = 105\ \beta^{3/2}$ below $P_0 = 0.316$ GV, $\lambda = 55\ \beta^{3/2}$ below 0.562 GV and $\lambda = 10.0$ g/cm$^2$ constant with rigidity below 1.0 GV. All of these new possibilities lead to a larger mean path length in g/cm$^2$ at lower energies and therefore more B production. These different dependences of the mean path length are illustrated in Figure 2.

The B/C ratio at low energies calculated using the above possibilities for the mean path length are shown in Figure 3. The calculated B/C ratio increases significantly with increasing path length. The value is 8% at 10 MeV/nuc for a path length ~10 g/cm$^2$. This is ~2 times the ratio calculated at 10 MeV/nuc for a break rigidity =1.0 GV where the path length is only 1.2 g/cm$^2$. This higher value is still >2.0 sigma below the measured value for the B/C ratio at this energy.

Note that the calculations all assume a truncation parameter 0.12. This modification of the PLD is necessary in order to fit the spectra of C, O, Ne, Mg, Si and Fe at low energies and also provides a good fit to the AMS measurements of the B/C ratio above 100 GeV/nuc (see Figure1). This truncation is described more fully in the following section.

**<u>B - Truncation of the Path Length Distribution</u>**

These calculations start by using the LBM (Webber and Higbie, 2009) in its strictest sense. This leads to an exponential distribution of path lengths at a mean path length $\lambda$ for all path lengths. In this connection we have shown in companion papers (Webber, 2015b, 2016) that the spectra of primary nuclei such as C, O, Ne, Mg, Si and Fe at energies below ~100 MeV/nuc (1.2 GV) do not fit this simple LBM representation but are better explained by the truncation effect. At energies >100 MeV/nuc, however, the pure LBM does indeed provide an excellent fit to the spectra of all of these primary nuclei (Webber, 2015b). The truncation used here is described in Webber, 2016, and follows the procedure described in Garcia Munoz, et al., 1987. In this calculation the value of the truncation parameter used to fit these primary nuclei spectra is 0.12. The non-exponential shape of the new PLD at given values of $\lambda$, for various values of the truncation parameter is shown in Figure 5 of Webber, 2016.

The B/C ratios calculated using a truncation parameter of 0.12 are shown in Figure 3. At low energies it predicts a B/C ratio between 4-8% at 10 MeV/nuc for the three path length variations used. This same truncation factor along with a continuation of the $P^{0.45}$ dependence of the mean path length at higher energies <u>is sufficient by itself to explain the observed B/C ratio above 100 GeV/nuc by AMS (Oliva, et al., AMS-CERN, 2015), however</u>. The maximum value of ~8% for the B/C ratio at 10 MeV/nuc is still >2.0 sigma below the measurement plus cross section uncertainties.

## **C - Including a Source Component of B as Produced in a Nested LBM**

Another possibility to increase the B/C ratio to still larger values, but still within the framework of a LBM, is that there is a "source" abundance of B itself. This could uniquely be caused by a component of B that is produced at or near the source itself, for example, by primary nuclei such as C and higher Z nuclei passing through matter concentrated near the source. This is sometimes called the Nested LBM (see e.g., Cowsik, Burch and Madziwa-Nussinov, 2014). If we assume that this local amount of matter is = 0.5 g/cm$^2$ (e.g., a thin target as opposed to a thick target for galactic propagation), then a B/C "source" component is produced with an abundance ratio ~3.5% x the C abundance. This abundance ratio is almost constant with energy above ~100 MeV/nuc.

The B/C ratio that is calculated using the standard LBM along with a 0.12 truncation parameter and including this Nested LBM "source" component is shown by the red curve in Figure 3. This additional component is significant because when it is combined with the simple LBM calculation with only ~1.2 g/cm$^2$ of interstellar material at 10 MeV/nuc, it produces a B/C ratio 9% at 10 MeV/nuc, about the same as produced by ~10 g/cm$^2$ of interstellar matter.

These nested LB source components would presumably be present for all other secondary nuclei, however, and would be difficult to reconcile with a wide variety of Voyager observations including the intensities of many secondary nuclei with Z=9-25.

## **Summary and Conclusions**

In this section we summarize the modifications to a simple LBM that will fit the B/C ratio from the lowest energies measured by Voyager 1 up to the highest energies measured by AMS-2. The simplest model that would fit both the low energy and high energy B/C data would be one in which most of the matter is traversed in the galaxy as in a LBM. This galactic path length distribution may be modified as a result of different values for the break rigidity, $P_0$. A "truncation" of path lengths may also be present for values of $\lambda \leq 4$ g/cm$^2$.

The mean path length at ~10 MeV/nuc, where a B/C ratio of 0.14 $\pm$ 0.03 is measured at Voyager, may therefore range from ~1.2-6 g/cm$^2$ which results in calculated B/C ratios which are between 4-8%. We find it difficult to produce B/C ratios at ~10 MeV/nuc which are $\geq 8\%$ in these LBM calculations.

However, in combination with a mean path length, $\lambda = 22.3\ \beta\ P^{-0.45}$ above 1 GV and $\lambda = 33.9\ \beta^{3/2}$ below 1 GV in the galaxy, the "truncation" approach alone will explain the AMS-2 measurements of the B/C ratio above 100 GeV/nuc without any further modification of the mean path length dependence on rigidity.

We also note here that the Nested LBM "source" component of B that is used in this paper to describe the B/C ratio is effectively an upper limit on the amount of matter that could have been traversed by primary nuclei such as, H and He and heavier nuclei, for example, near the source, without producing observable effects on the abundance of various low energy secondary cosmic ray nuclei. If we consider this upper limit to be ~1 g/cm$^2$ at low energies and $\leq 0.5$ g/cm$^2$ at high energies, then we note that this value is much smaller than the amount of matter near the source that would need to have been traversed by cosmic ray H and He nuclei in order to produce the positron fraction ~0.15 at ~300 GeV/nuc that is also observed by AMS-2 (Kounine, et al., AMS-CERN, 2015). In the calculations of Cowsik, Burch and Madziwa-Nussinov, 2014, using a Nested LBM that could produce the observed positron fraction, most of the secondary production of positrons would, in fact, occur near the sources and not in the galaxy in general. So different scenarios would be needed to explain both the high energy positron fraction and the low energy B/C ratio.

**Acknowledgements:** The authors are grateful to the Voyager team that designed and built the CRS experiment with the hope that one day it would measure the galactic spectra of nuclei and electrons. This includes the present team with Ed Stone as PI, Alan Cummings, Nand Lal and Bryant Heikkila, and to others who are no longer members of the team, F.B. McDonald and R.E.

Vogt. Their prescience will not be forgotten. This work has been supported throughout the more than 35 years since the launch of the Voyagers by the JPL.

# References

Cowsik, R., Burch, B. and Madziwa-Nussinov, T., 2014, Ap.J. 786, 7

Cummings, A.C., Stone, E.C., Lal, N., Heikkila, B. and Webber, W.R., 2016, Ap.J., submitted

Garcia-Munoz, M., Simpson, J.A., Guzik, T.G., Wefel, J.P. and Margolis, S.G., 1987, Ap.J. Supple Series, 64, 269

Kounine, A., et al., AMS-CERN, 2015

Kraichnan, R. H., 1965, Physics of Fluids, 8, 1385

Lave, K. A., Wiedenbeck, M.E., Binns, W.R., et al. 2013, Ap. J., 770, 117

Oliva, A., et al., AMS-CERN, 2015

Shapiro, M.M. and Silberberg, R., 1970, Ann. Rev. Nucl. Sci., 20, 323

Webber, W.R., 1993, Ap.J., 402, 188-194

Webber, W.R., Higbie, P.R. and McCracken, K.G., 2007, JGR, 112, A10106

Webber, W.R. and Higbie, P.R., 2009, JGR, 114, A02103

Webber, W.R. and Higbie, P.R., 2013, http://arXiv.org/abs/1308.6598

Webber, W.R., McDonald, F.B. and Lukasiak, A., 2003, Ap.J., 559, 582-595

Webber, W.R., 2015a, http://arXiv.org/abs/1508.01542

Webber, W.R., 2015b, http://arXiv.org/abs/1508.06237

Webber, W.R., 2016, http://arXiv.org/abs/1604.06477

**FIGURE CAPTIONS**

**Figure 1:** The B/C ratio as a function of energy. Shown in the figure are the new Voyager measurements (Cummings, et al., 2016), and the new AMS-2 measurements (Oliva, et al., CERN, 2015). Also included are previous Voyager measurements in the inner heliosphere (Webber, McDonald and Lukasiak, 2003) open circles and ACE measurements at the Earth (Lave, et al., 2013) solid circles both at a solar modulation level $\phi=250$ MV. The calculations are for the standard LBM with $\lambda = 22.3 \beta \; P^{-0.45}$ for $P > 1.0$ GV and $\lambda = 33.9 \beta^{3/2}$ for $P < 1.0$ GV (black line) and this standard model with the addition of a truncation = 0.12 at short path lengths (dashed black line) at high and low energies. Also shown at low energies are the calculated B/C ratios for solar modulation parameters, $\phi = 250$, 500 and 1000 MV. These fit the observations of AMS-2, Voyager and ACE.

**Figure 2:** Path lengths in g/cm$^2$ that are used in LBM. The examples used here are for $P_0=0.316$ and 1.0 GV. Below $P_0$ the path length is assumed to be ~$\beta^{3/2}$. A path length = 10.7 g/cm$^2$ is shown as a dashed line below ~120 MeV/nuc.

**Figure 3:** The B/C ratio as a function of energy below 1 GeV/nuc. The data are the same as shown in Figure 1. The calculations are shown for path lengths with $P_0 = 0.316$ and 1.0 GV and a path length = 10.7 g/cm$^2$ below 120 MeV/nuc as shown in Figure 2 and also for a Nested LBM as described in the text (in red). The red line shows the B/C ratio for 1.0 g/cm$^2$ of material near the source along with a value of $P_0 = 1.0$ GV for galactic propagation.

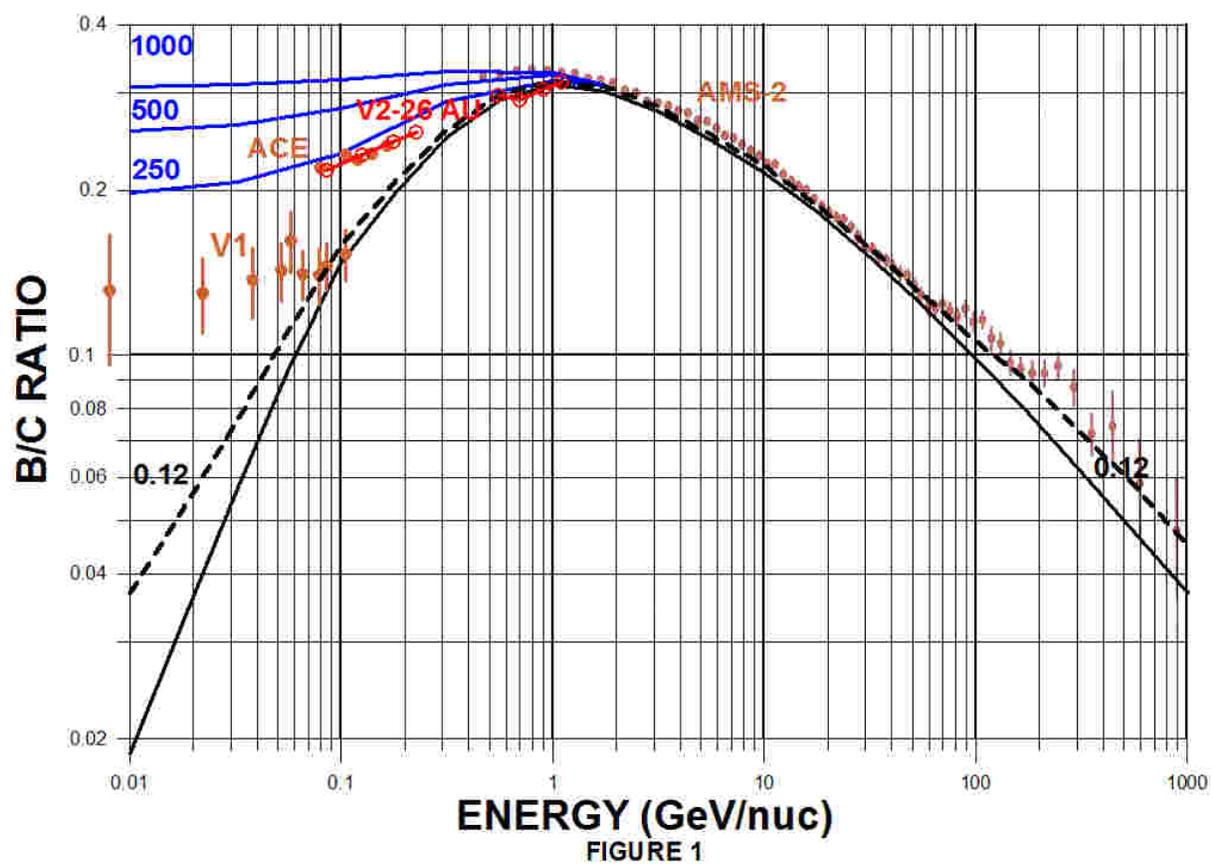

FIGURE 1

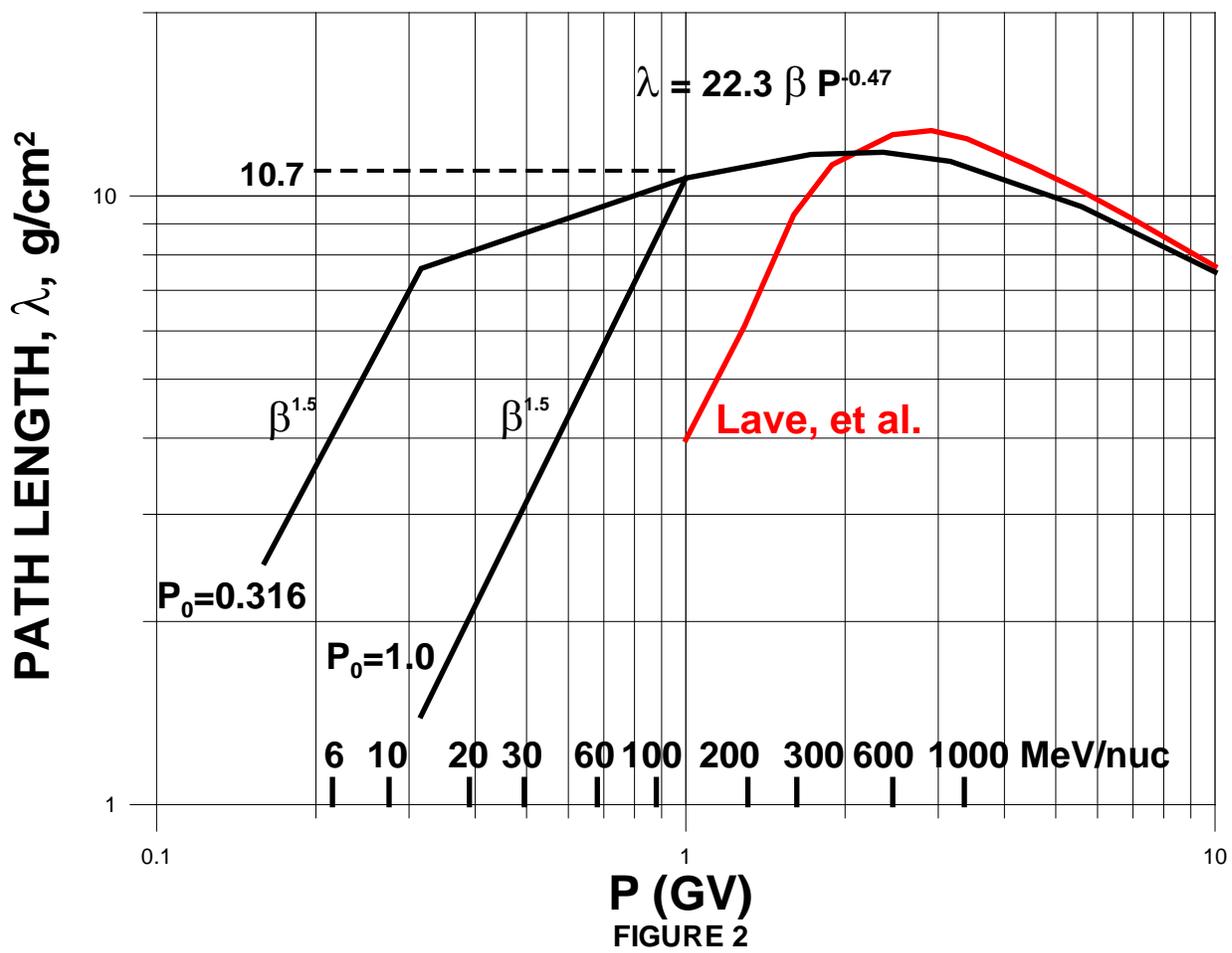

FIGURE 2

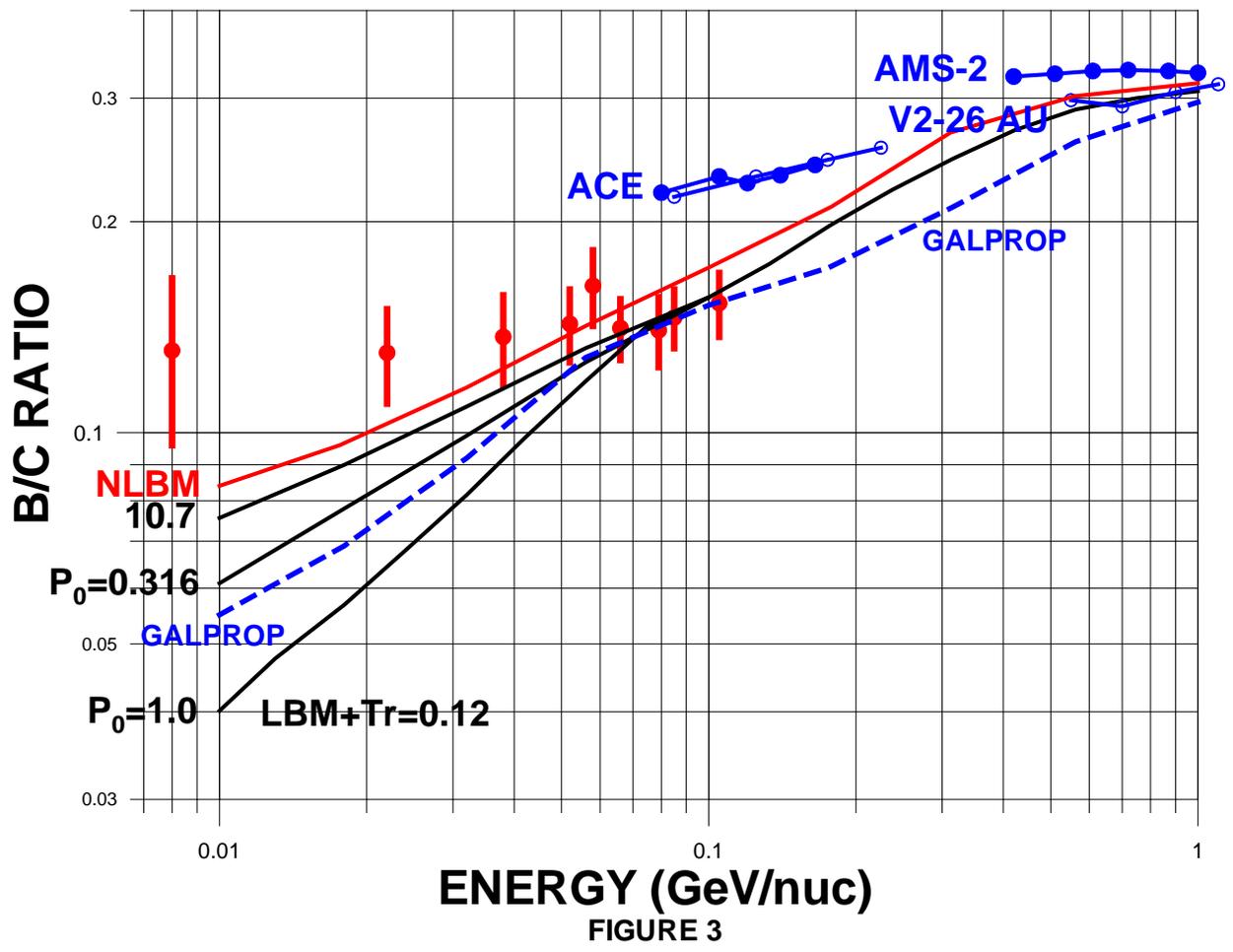

FIGURE 3